\documentclass[conference]{IEEEtran}
\IEEEoverridecommandlockouts
\usepackage{cite}
\usepackage{amsmath,amssymb,amsfonts}
\usepackage{algorithmic}
\usepackage{graphicx}
\usepackage{textcomp}
\usepackage{xcolor}

\newcommand{\Nf}{$N_\mathrm{filter}$}

\ifCLASSOPTIONcompsoc
    \usepackage[caption=false, font=normalsize, labelfont=sf, textfont=sf]{subfig}
\else
\usepackage[caption=false, font=footnotesize]{subfig}
\fi

\def\BibTeX{{\rm B\kern-.05em{\sc i\kern-.025em b}\kern-.08em
    T\kern-.1667em\lower.7ex\hbox{E}\kern-.125emX}}
\begin{document}

\title{Attention Based Neural Networks for Wireless Channel Estimation}

\author{\IEEEauthorblockN{Dianxin Luan, John Thompson}
\IEEEauthorblockA{\textit{Institute for Digital Communications, School of Engineering, University of Edinburgh, Edinburgh, EH9 3JL, UK}\\
Email address : Dianxin.Luan@ed.ac.uk, john.thompson@ed.ac.uk}
}

\maketitle

\begin{abstract}
In this paper, we deploy the self-attention mechanism to achieve improved channel estimation for orthogonal frequency-division multiplexing waveforms in the downlink. Specifically, we propose a new hybrid encoder-decoder structure (called HA02) for the first time which exploits the attention mechanism to focus on the most important input information. In particular, we implement a transformer encoder block as the encoder to achieve the sparsity in the input features and a residual neural network as the decoder respectively, inspired by the success of the attention mechanism. Using 3GPP channel models, our simulations show superior estimation performance compared with other candidate neural network methods for channel estimation. 
\end{abstract}


\section{Introduction}
For fifth generation (5G) communication systems, orthogonal frequency-division multiplexing (OFDM) demodulation needs precise channel state information in order to compensate for the distortion of the channel and provide robust communication. Conventionally, the least-squares (LS) and minimum mean squared error (MMSE) methods are used for channel estimation. However, the estimates from the LS method are excessively noisy and the practical implementation of the MMSE method is difficult as the accurate channel statistics cannot be accessed in advance. This motivates us to explore machine learning solutions for channel estimation in complex propagation environments. 

Compared with the conventional methods aiming to find an analytical solution, neural network methods can provide improved performance. Some recent investigations into the channel estimation problem has led to neural networks being proposed, such as ChannelNet \cite{b2}, ReEsNet \cite{b3} and Interpolation-ResNet (called \textsl{InterpolateNet} in this paper) \cite{b4}. ChannelNet \cite{b2} is one of the first deep learning algorithms proposed for channel estimation in communication systems by combining SRCNN \cite{b5} and DnCNN \cite{bI}. However, conventional deep learning algorithms may suffer from the degradation problem, which inspires the use of a residual network architecture to solve this. ReEsNet \cite{b3} is a residual convolutional neural network that can outperform ChannelNet \cite{b2} with a reduced number of learned parameters. In addition, when compared with ReEsNet, the recently proposed InterpolateNet \cite{b4} can achieve a slightly improved estimation performance and the number of learned parameters is further reduced by 82\%. The attention mechanism has also been studied recently. The paper \cite{bA} introduces a non-local attention \cite{bB} to implement the channel estimation for a multiple antenna OFDM system. A graph attention network has also been utilized for channel estimation \cite{bC}. The papers \cite{bD} \cite{bE} \cite{bF} also deploy the attention mechanism to improve the performance. 

For the channel estimation neural networks, LS is widely exploited as the input. However, conventional neural network solutions process the input features equally. It practices the importance of each channel gain is not necessarily equal for all the subcarriers due to the fact that the correlation matrix of channel gains will not be diagonal. It indicates that some elements of LS estimation may be more significant than others, and the neural network may need to focus on an important subset of the LS estimates. To solve that, the transformer encoder involving multi-head attention \cite{b12} attracts our interest. It helps capture the useful information/features (the elements of LS estimate with a strong correlation to channel predictions) and is expected to focus on different parts in different heads. Unlike conventional neural networks that process all the LS estimates equally, a pre-processing architecture is needed to focus on the most important elements among the LS estimate. This will ensure that the subsequent neural network can focus on the critical features to yield the best channel estimate. 
%
%

In this paper, we propose for the first time a novel hybrid encoder-decoder solution called HA02 for channel estimation in the downlink scenario. It deploys a transformer encoder block based on \cite{b12} and a residual neural network for the decoder. Instead of processing all the input features equally, the encoder processes the original input features to explore the importance of each one, to ensure the decoder to focus on the significant features. From the simulation results, HA02 can achieve the best performance for channel estimation among comparable neural network methods. The proposed solution should also be robust to the extended channel conditions, such as signal-to-noise ratio (SNR) and Doppler shift. We also implement a transformer architecture from \cite{bD} for comparison, whose performance is found to be inferior to HA02. 

Section \ref{System Architecture} presents the baseband and frame structure which is based on the 5G New Radio (NR) standard \cite{b7}. Section \ref{Conventional methods for channel estimation} introduces several methods for channel estimation and describes the Channelnet, ReEsNet and InterpolateNet networks. Section \ref{Hybrid architecture HA02} proposes the system description of HA02. Section \ref{Simulation results} presents key simulation results comparing the different methods. Section \ref{Conclusion} summarizes the key findings of the paper. 
\section{System Architecture}
\label{System Architecture}
\subsection{Baseband architecture}
\label{Baseband Architecture}
OFDM is deployed as the baseband modulator. At the transmitter, the bit level signal $s(i)$ is processed in a Quadrature Phase Shift Keying (QPSK) modulator with Gray coding and then pilot signals are inserted. Each slot consists of $N_s = 14$ OFDM symbols and each OFDM symbol $N_f = 72$ subcarriers, which matches 6 5G NR resource blocks. The 1\textsuperscript{st} and 13\textsuperscript{th} OFDM symbols are reserved for pilots ($N_{pilot} = 2$) \cite{b7}. For the first pilot OFDM symbol, the indices of the pilot subcarriers start from subcarrier 1 and are spaced by 2 subcarriers. For the second pilot symbol, the pilot subcarriers start at subcarrier 2 and are spaced by 2 subcarriers. The remaining subcarriers in the pilot symbols are set to 0. All data subcarriers in the remaining 12 OFDM symbols are assigned to QPSK-modulated symbols. The Inverse fast Fourier transform (IFFT) converts the frequency domain data symbols to the time domain OFDM signal samples where $T_{Sym}$ denotes the duration of each pilot-data symbol. Then, the Cyclic Prefix (CP of duration $T_{CP}$) is added to the front of each symbol to provide resistance to multipath effects. Normally the FFT and IFFT operators use scaling factors of 1 and $\left(1/{N_f}\right)$, but those are changed here to $\left(1/\sqrt{N_f}\right)$ to avoid changing the power of the FFT/IFFT outputs. 
%
%
The channel is assumed to be a multipath fading channel with $M$ paths and the corresponding impulse response is defined by 
\begin{equation}
h(t) = \sum_{m=0}^{M-1} a_m\delta(t-\tau_m)
\end{equation}
Where $a_m$ is the magnitude of the $m^{th}$ path, $\tau_m$ is the corresponding delay with the condition that $\tau_m \in [0, T_{CP}]$. The received signal $y(n)$ in the time domain is given by 
\begin{equation}
y(n) = h(n)\otimes x(n) + w(n)
\end{equation}
Where $\otimes$ denotes the convolution operation, $w(n)$ represents the additive white Gaussian noise (AWGN) and $x(n)$ is the transmitted signal in the time domain. The CP is removed and the received data is converted into the frequency domain using the FFT operation. The received signal at the $k^{th}$ subcarrier and $i^{th}$ OFDM symbol, $Y(k, i)$, is represented by 
\begin{equation}
Y(k, i) = H(k, i)X(k, i) + W(k, i)
\end{equation}
Where $H(k, i)$, $X(k, i)$ and $W(k, i)$ are the Discrete Fourier Transforms (DFT) of $h(n)$, $x(n)$ and $w(n)$ at the $k^{th}$ subcarrier and the $i^{th}$ OFDM symbol. The received pilot signal is extracted to provide a channel amplitude and phase reference for the data symbols in the complete packet. Then the recovered data symbols are processed in a QPSK demodulator to estimate the received bit-level data $\hat{s}(i)$. 
\subsection{Channel model and frame structure}
We consider a single-input-single-output (SISO) and Rayleigh fading scenario modelled by the Generalized Method of Exact Doppler Spread \cite{b8}. The Extended Typical Urban (ETU) defined in 3GPP TS 36.101, representing a high delay spread environment, is deployed for simulation. 
The system is assumed to operate at 2.1GHz (sub-6GHz band) and the subcarrier spacing is $15$kHz. Each frame consists of 10 slots. The corresponding slot duration is $1$ms and each slot corresponds to a single channel realization $H$. Each OFDM symbol duration is $66.7$us (sampling rate is $1.08$MHz) and the CP length is used to be 16 OFDM samples ($14.81$us). 
\section{Conventional methods and neural networks for channel estimation}
\label{Conventional methods for channel estimation}
The conventional channel estimation methods are the LS and MMSE techniques used to compare the performance with the neural network solutions. The algorithm implementations of the LS and MMSE methods are explained and the conventional neural network methods are also introduced. 
\subsection{LS method}
By minimizing the mean squared error (MSE) between $Y$ and $H \circ X$ at the pilot positions, i.e. $\mathop{\arg\!\min_{H}}\Arrowvert Y-H \circ X\Arrowvert^{2}_{2}$ at the pilot positions, to give an estimate of $H$, the frequency domain LS estimation is given by 
\begin{equation}
    \hat{H}_{LS} = \frac{Y_{Pilot}}{X_{Pilot}}
\end{equation}
Where $Y_{Pilot}, X_{Pilot} \in \mathbb{C}^{\frac{N_f}{2}\times N_{pilot}}$ denotes the received and transmitted pilot signals respectively and the mathematical division operation is performed elementwise. The dimensions are consistent with the definition of the pilot signals in Section.~\ref{Baseband Architecture}. LS is easy to implement with extremely low complexity. The prediction $\hat{H}_{LS}$ is then resized by bilinear interpolation \cite{b15} in both the time and frequency domain to estimate the complete channel matrix of size $\mathbb{C}^{N_f \times N_s}$. 
\subsection{Frequency Domain MMSE method (FD-MMSE)}
To achieve the minimum distance between $H$ and $H_{LS}$, i.e. $\mathop{\arg\!\min_{H}}\Arrowvert H - \hat{H}_{LS}\Arrowvert^{2}_{2}$, the frequency domain linear MMSE method of the $j^{th}$ OFDM symbol $\hat{H}_{MMSE}(j)$ is \cite{b11} 
\begin{equation}
\label{MMSE}
    \hat{H}_{MMSE}(j) = R_{HH_{p}}(j)\left(R_{H_{p}H_{p}}(j) + \frac{\sigma_N^2}{\sigma_X^2}I\right)^{-1}\hat{H}_{LS}(j)
\end{equation}
Where $i$ refers to the index of the OFDM symbols for pilot in the slot with the restriction that $i = 1, 13$ and $j = 1, 2$ denotes the index of the pilot symbol. $H(i) \in \mathbb{C}^{N_f}$ is the channel gain vector for the $i^{th}$ OFDM symbol and $H_p(i) \in \mathbb{C}^{\frac{N_f}{2}}$ denotes the channel matrix for the $i^{th}$ OFDM symbol for pilot and the corresponding pilot subcarriers. (${\sigma_N^2}/{\sigma_X^2}$) is the numerical reciprocal of the SNR and $I$ is the corresponding identity matrix. The matrices $R_{HH_p}(j) = E\{H(i)H_{p}(i)^{H}\}$ and $R_{H_pH_p}(j) = E\{H_{p}(i)H_{p}(i)^{H}\}$ are the correlation matrix. Therefore, $\hat{H}_{MMSE}$ has a size of $\mathbb{C}^{N_f \times N_{pilot}}$. To reduce complexity, bilinear interpolation is also implemented in time to predict the remaining channel gains in one slot. By utilizing prior statistical knowledge of the channel state, the FD-MMSE method improves the performance of LS method, but requires non-causal statistical information. Therefore, physical implementation of the FD-MMSE method is challenging. 
\subsection{ChannelNet, ReEsNet and InterpolateNet}
ChannelNet \cite{b2} (670,000 learned parameters) is a cascaded neural network designed for channel estimation, which combines a super-resolution block and an image-restoration module. ReEsNet \cite{b3} is a residual convolutional neural network with around 53,000 parameters, which outperforms ChannelNet. InterpolateNet \cite{b4} (9,442 learned parameters) can achieve an almost same performance and 82\% reduced complexity compared with ReEsNet. As other neural network solutions are much less complex than ChannelNet, we only consider the ReEsNet and InterpolateNet for simulation. 
\section{Hybrid architecture: HA02}
\label{Hybrid architecture HA02}
\begin{figure}[htbp]
\centerline{\includegraphics[width=0.5\textwidth]{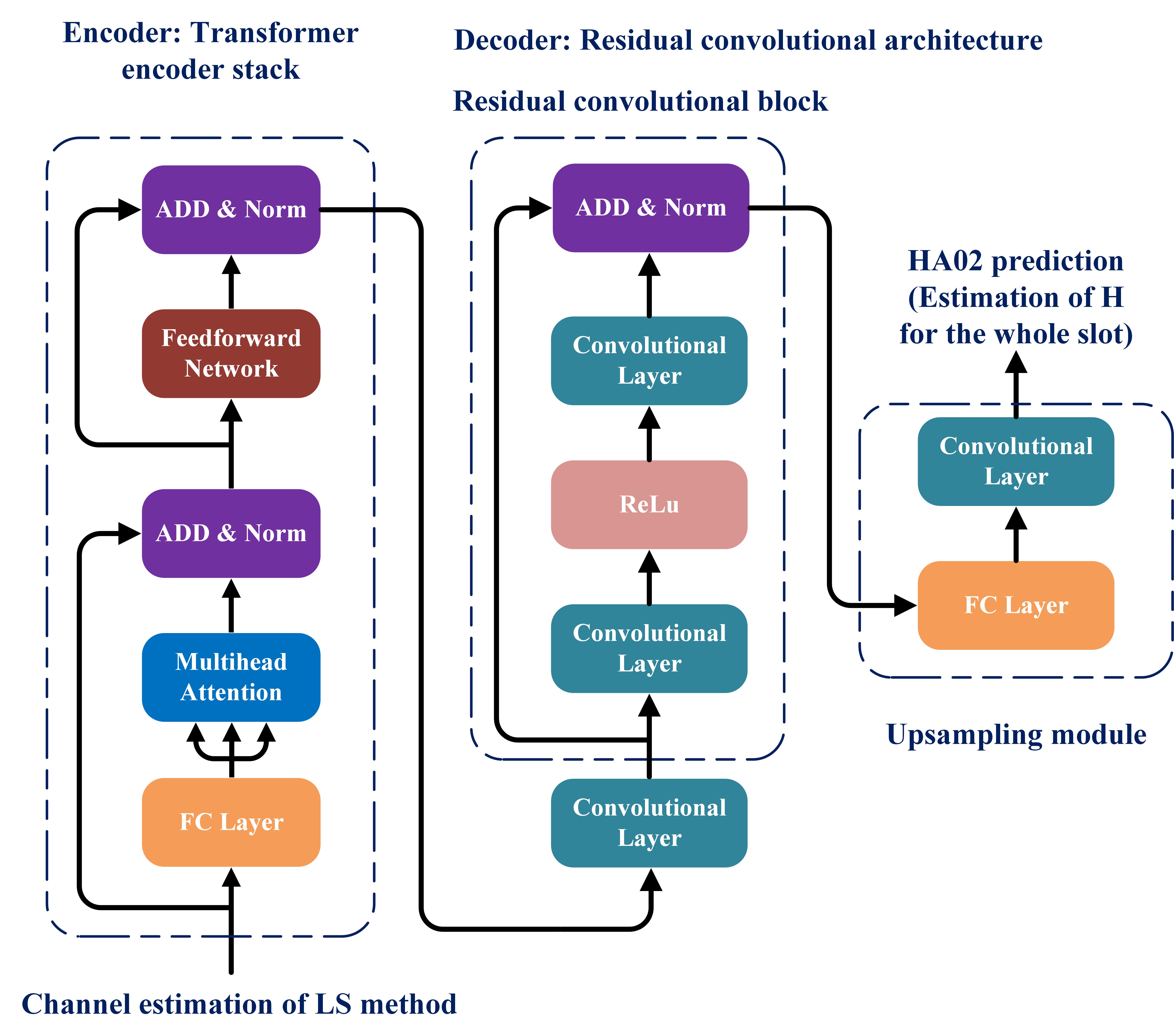}}
\caption{Hybrid architecture HA02}
\label{HA02}
\end{figure}

In this paper, our work has considered the multi-head attention mechanism \cite{b12}. The aim is to explore the possibility of enhancing the overall performance by achieving the sparsity among the input features. For the input of the hybrid architecture HA02, the matrix $\hat{H}_{LS} \in \mathbb{C}^{\left(\frac{N_f}{2}\right) \times N_{pilot}}$ for all pilot symbols is concatenated to be one column vector $\in \mathbb{C}^{\frac{N_{pilot}N_f}{2}}$. The real and imaginary parts of the one column vector are split into two different channels as the second dimension of the input. Therefore, the input of HA02 has a size of $\mathbb{R}^{\left(\frac{N_{pilot}N_f}{2}\right) \times 2}$. The model is trained by the channel matrix of the complete slot to achieve the interpolation in both frequency dimension and time dimension. Therefore, the output of HA02 is of size of $\mathbb{R}^{(N_{s}{N_f}) \times 2}$, where the index 1 of the second dimension is the real part and the index 2 is the imaginary part. 
The corresponding complex channel estimate, denoted as $\hat{H}_{\mathrm{HA02}}$, is of size of $\mathbb{C}^{{N_f} \times N_{s}}$. The hybrid architecture shown in Fig.~\ref{HA02} involves two substructures (encoder and decoder), which are the transformer encoder stack and the residual convolutional architecture. 
\subsection{Encoder: Transformer encoder stack}
\label{Encoder: Transformer stack}
In this paper, we implement a transformer encoder block following the original paper \cite{b12}, which involves multi-head attention and a feed-forward network. 
\subsubsection{\textbf{Attention module and normalization}}
The encoder of transformer is based on the self-attention mechanism. All the inputs to the multi-head attention model are generated from the input after a linear transformation. A fully connected layer is used for this, which resizes the input from $\mathbb{R}^{\left(\frac{N_{pilot}N_f}{2}\right) \times 2}$ to $\mathbb{R}^{\left(\frac{3N_{pilot}N_f}{2}\right) \times 2}$, to be the input of the multi-head attention module \cite{b12}. It splits that input to 3 sub-inputs, which are named the key, the query and the value $K, Q, V \in \mathbb{R}^{\left(\frac{N_f}{2}\right) \times 2 \times N_{head}}$. Here $N_{heads}$=$N_{pilot}$ is the head size. Moreover, we do not apply any mask in the transformer encoder block \cite{b12}. After scaled dot-product attention \cite{b12} is computed by equ.~(\ref{Attentions}) and concatenating the scaled dot-product attentions from different heads in the first dimension, the concatenated result is forwarded to the second fully connected layer to generate the output of the attention module. 
\begin{equation}
    \mathrm{Attention} = \mathrm{softmax}\left(\frac{QK^{T}}{\sqrt{\frac{N_f}{2}}}\right)V
\label{Attentions}
\end{equation}
The Add \& Norm layer adds the output from the attention module and the skip connection from the first fully-connected layer input, and applies layer normalization \cite{layer} to the superimposed result. The output of the Add \& Norm layer $\in \mathbb{R}^{\left(\frac{N_{head}N_f}{2}\right) \times 2}$ is then processed by the feed-forward network and normalization module. 
\subsubsection{\textbf{Feed-forward network and normalization}}
Identical to \cite{b12}, the feed-forward network involves three layers, which are one fully connected layer, one activation layer and another fully connected layer connected in series. GeLu \cite{b13} is deployed as the activation function for the encoder. The Add \& Norm layer adds the output from the feed-forward network and the skip connection from the input, and applies the layer normalization \cite{layer} to the superimposed result. The normalized result is the input $\in \mathbb{R}^{\left(\frac{N_{head}N_f}{2}\right) \times 2 \times 1}$ to the decoder. 
\subsection{Decoder: Residual convolutional architecture}
The implemented residual convolutional architecture shown in Fig.~\ref{HA02} involves 7 layers, because the residual architecture can easily solve the degradation problem. It is composed of one convolutional layer, one residual convolutional module which consists of a stack of 1 residual block (By using residual block with skip connection, the residual convolutional block can contain a more large number of residual block without the impact of degradation problem and is more powerful for the channel estimation task. We set that to 1 for the complexity reduction and from the simulations it has already outperformed the other methods), and one upsampling module in series. The first convolutional layer has \Nf \ filters with a kernel size of {2 $\times$ 2 $\times$ 1} followed by a residual convolutional block. The residual block consists of one convolutional layer with \Nf \ filters, each corresponding to a kernel size of {2 $\times$ 2 $\times$ \Nf}, followed by one ReLu layer and one convolutional layer with \Nf \ filters. The kernel size of each is {2 $\times$ 2 $\times$ \Nf}. The Add \& Norm layer processes the superimposed result of the input and output of the residual block to the upsampling module. The upsampling module consists of one fully connected layer and one convolutional layer. To achieve one-dimensional upsampling for the first dimension and improve the generalization to the unknown SNR and Doppler shift \cite{b14}, we deploy the fully connected layer in the residual convolutional architecture. The fully connected layer resizes the output of residual block from $\mathbb{R}^{\frac{N_{pilot}N_f}{2} \times 2 \times N_{filter}}$ to the size of $\mathbb{R}^{\left(N_{s}{N_f}\right) \times 2 \times N_{filter}}$. The last coupled convolutional layer has 1 filter with the kernel size of {2 $\times$ 2 $\times$ \Nf} to generate the output $\in \mathbb{R}^{N_{s}{N_f} \times 2}$. We use \Nf \ equal to 2, therefore, the total number of parameters for HA02 is 105,607 (99.54\% are contributed by fully-connected layers): 31,824 for the encoder and 73,783 for the decoder. 
\subsection{Transformer-based method (TR) for comparison}
We have also implemented a transformer-based method similar to \cite{bD} with 31,829 parameters for comparison. The output of TR is interpolated by the bilinear method to estimate the channel matrix of the whole slot. 
\section{Simulation results}
\label{Simulation results}
MSE is a key performance metric to evaluate the error between the desired and estimated channels, defined as 
\begin{equation}
    \mathrm{MSE}(\hat{H}, H) = \frac{1}{N_f N_s}\sum_{i=1}^{N_f} \sum_{j=1}^{N_s} {\left\Arrowvert\hat{H}_{ij} - H_{ij}\right\Arrowvert_{2}^{2}}
\end{equation}
Where $H_{ij}$ is the real channel at subcarrier $i$ and OFDM symbol $j$ and $\hat{H}_{ij}$ is the corresponding estimate. The denoising gain $G=10\log_{10}\frac{\Arrowvert \hat{H}_{LS}-H\Arrowvert_{2}^{2}}{\Arrowvert \hat{H}_{\mathrm{HA02}}-H\Arrowvert_{2}^{2}}$ is also measured, but the definition is modified from \cite{opt} to ensure a positive gain value. The training dataset is generated on the ETU channel (SNR from 5dB to 25dB and maximum Doppler shift from 0Hz to 97Hz) and composed of 95,000 samples, 95\% for training and 5\% for validation. In this paper, all of the neural networks deployed are trained using the training dataset generated by the same channel. The hyperparameters parameters for training the neural networks are described in Table.~\ref{Training Parameter}. 
\begin{table}[htbp]
\caption{Training Parameters}
\begin{center}
\begin{tabular}{|c|c|c|c|}
\hline
\textbf{}& \textbf{HA02\&TR}& \textbf{ReEsNet}& \textbf{InterpolateNet}\\
\hline
\textbf{Reference} & This Paper & \cite{b3} & \cite{b4} \\
\hline
\textbf{Optimizer}& Adam& Adam& Adam\\
\hline
\textbf{Maximum epoch}& 100 & 100& 100\\
\hline
\textbf{Initial learning rate (lr)}& 0.002& 0.001& 0.001\\
\hline
\textbf{Drop period for lr}& every 20& None& every 20\\
\hline
\textbf{Drop factor for lr}& 0.5& None& 0.5\\
\hline
\textbf{Minibatch size}& 128& 128& 128\\
\hline
\textbf{L2 regularization}& 1e-7& 1e-7& 1e-7\\
\hline
\end{tabular}
\label{Training Parameter}
\end{center}
\end{table}

The loss function deployed for training the ReEsNet and InterpolateNet networks is MSE \cite{b3} \cite{b4}. The loss function for HA02 and TR is the Huber loss defined in Equation.~(\ref{huber}) with transition point $\delta$ of 1, which can lessen the impact of outliers. 
\begin{equation}
 L_{\delta}(a) =
\begin{cases}
\frac{1}{2}a^2& \text{if $|a| \leq \delta$}\\
\delta(|a| - \frac{1}{2}\delta)& \text{otherwise}
\end{cases}
\label{huber}
\end{equation}
\subsection{MSE performance over the extended SNR range}
\label{MSE performance SNR}
The test dataset is generated on the ETU channel with the extended SNR range from 0dB to 30dB and maximum Doppler shift from 0Hz to 97Hz (Mobile speed from 0km/h to 50km/h). For each method, each SNR is tested with 5000 independent channel realizations to average out Monte Carlo effects. 
\begin{figure}[htbp]
\centerline{\includegraphics[width=0.37\textwidth]{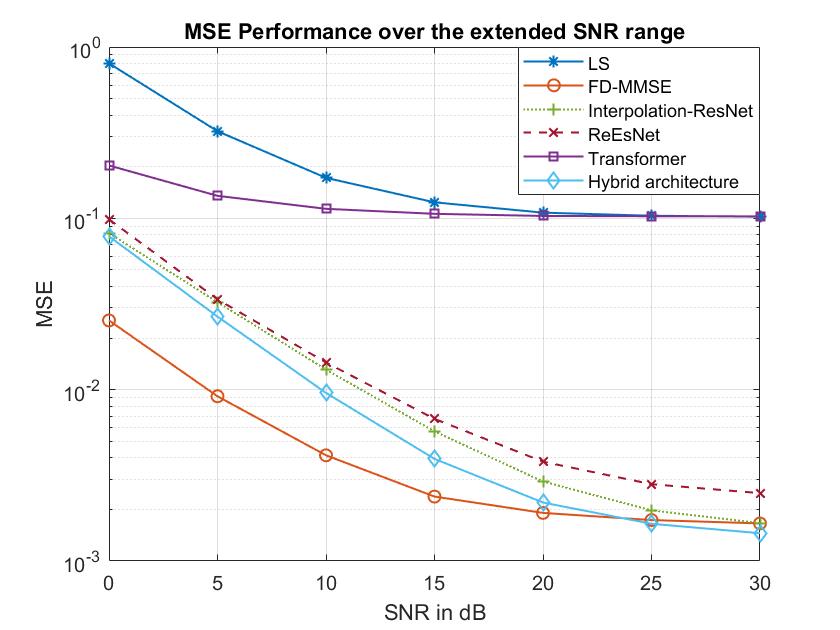}}
\caption{MSE performance over the extended SNR range}
\label{MSE performance over SNR (ETU)}
\end{figure}
Fig.~\ref{MSE performance over SNR (ETU)} compares the MSE results of each method when tested on the ETU channel. HA02 outperforms the LS method, ReEsNet, InterpolateNet and TR for all SNRs by applying the attention mechanism. At the SNR range from 0dB to 25dB, the FD-MMSE method outperforms HA02. At the high SNR range from 25dB to 30dB, HA02 achieves superior performance compared with the FD-MMSE method, because the FD-MMSE method only has access to the real channel matrix of the pilot symbols while the neural network solutions are trained by the real and complete channel matrix. 
\subsection{MSE performance over the extended Doppler shift range}
The test dataset is also generated on the ETU channel, with an extended maximum Doppler shift from 0Hz to 194Hz (Mobile speed from 0km/h to 100km/h) and a specific SNR=10dB. Each maximum Doppler shift is tested with 5000 independent channel realizations to average out Monte Carlo effects for each method. 
\begin{figure}[htbp]
\centerline{\includegraphics[width=0.37\textwidth]{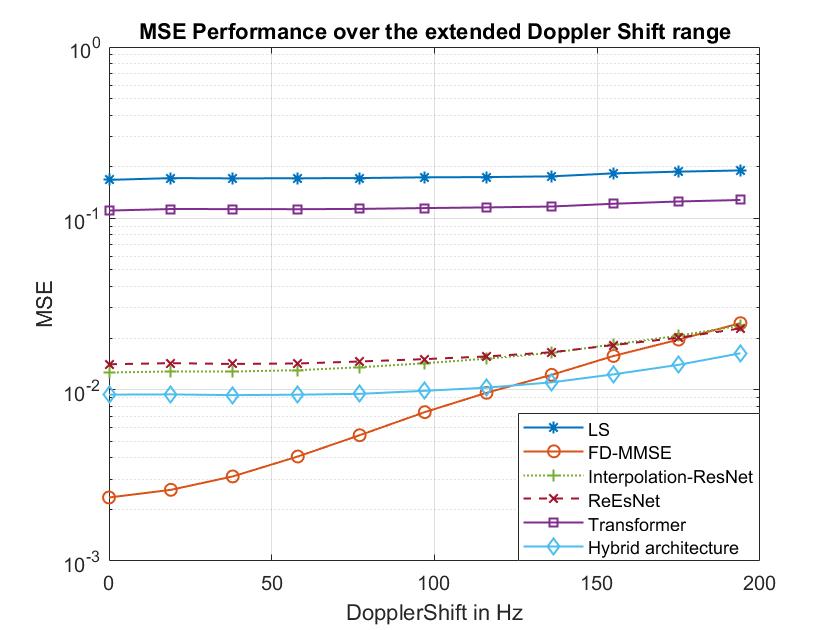}}
\caption{MSE performance over the extended Doppler shift range}
\label{MSE performance over Doppler shift (ETU)}
\end{figure}
Fig.~\ref{MSE performance over Doppler shift (ETU)} compares the MSE results of each method when tested on the ETU channel. HA02 also significantly outperforms LS, ReEsNet, InterpolateNet and TR methods. It is worse than the FD-MMSE method in the low Doppler shift scenarios. When the Doppler shift is low, the variation of the channel matrix for each symbol (in the time dimension) is negligible. It indicates that bilinear interpolation is comparable with the interpolation achieved by the neural networks. However, at the high Doppler range of 139Hz to 197Hz, HA02 has superior performance than the FD-MMSE method. Due to the significant impact derived by the Doppler shift in the time dimension, bilinear interpolation is much worse than the interpolation achieved by neural networks, as the neural networks are trained by the complete channel matrix to optimize both denoising and interpolation. 
\subsection{Denoising gain and complexity reduction}
Weight-level pruning method deployed, which reduces the computational complexity of neural networks by removing the redundant neural connections without retraining followed, is applied to the proposed method to reduce the required computation. 10\% pruning means that 10\% of the most insignificant weights are set to zero for both encoder and decoder. 
\begin{figure}[htbp]
\centerline{\includegraphics[width=0.37\textwidth]{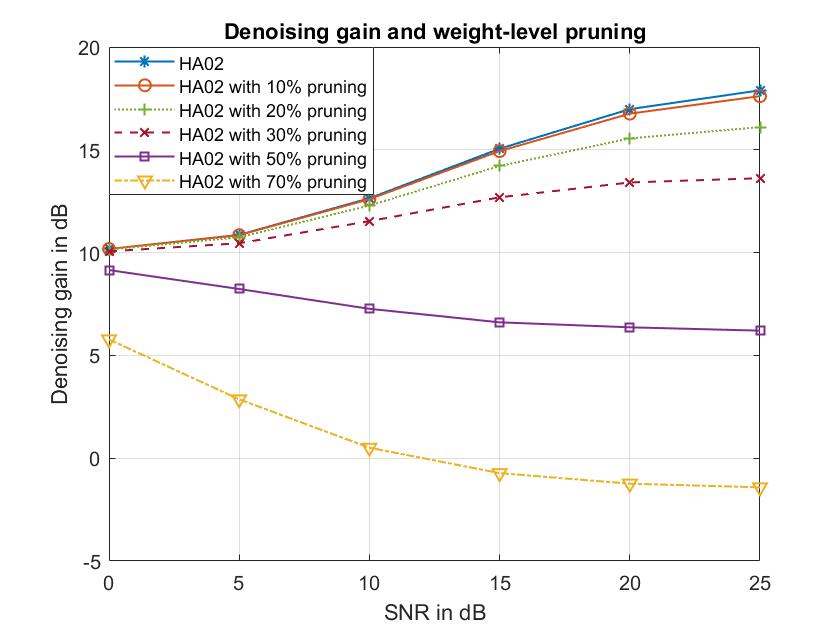}}
\caption{Denoising gain over the extended SNR range}
\label{Denoising gain over SNR}
\end{figure}
Fig.~\ref{Denoising gain over SNR} compares the denoising gains of different pruning ratios over the extended SNR range. 10\% pruning (95,046 parameters retained) is a preferred choice for that pruning as it leads only to a minimal reduction in denoising gain, less than 1dB. A pruning ratio not exceeding 30\% (73,924 parameters) leads to a consistent denoising gain of 10 dB across SNR values for the proposed model. However, 50\% pruning starts to degrade the denoising performance significantly which actually reduces to about 6dB when increasing the SNR. 
\section{Conclusion}
\label{Conclusion}
We proposed a hybrid neural network architecture HA02 that combines the transformer encoder block and the residual convolutional neural network for channel estimation in the downlink scenario. The encoder emphasises the critical features for the decoder to process. Compared with other neural network solutions, the performance of HA02 is improved by focusing on these important features. From the simulation results, HA02 reduces the MSE by 70\% for the SNR range from 15dB to 30dB, compared with ReEsNet. HA02 also can generalize better to extended SNR and Doppler shift settings. 
\section{Acknowledgement}
The authors gratefully acknowledge the funding of this research by Huawei. 

\vspace{12pt}

\end{document}